\documentstyle[epsf]{myl-aa} %for non-referee version

\begin{document}

   \thesaurus{     % A&A Section 6: Form. struct. and evolut. of stars
              (09.03.2)  % cosmic rays,
             } 
   
%
%a   \title{Cherenkov Light based Measurement of Extensive Air Showers
%a         to Determine the Energy Spectrum
%a         and Composition of Charged Cosmic Rays 
%a         between 0.3 and 10\,PeV}
\title {Energy Spectrum and Chemical
Composition of Cosmic Rays between 0.3 and 10 PeV
determined from the Cherenkov-Light and Charged-Particle distributions in 
Air Showers}

%a   \titlerunning{Energy Spectrum and Composition of Charged Cosmic Rays}
 
%   \subtitle{I. Comparison of Air Shower Measurements 
%               and Simulations}

   \author{The HEGRA-Collaboration: \\
F. Arqueros\inst{3},
J.A. Barrio\inst{2,3},
K. Bernl\"ohr\inst{1,7},
H. Bojahr\inst{6},
I. Calle\inst{3},
J.L. Contreras\inst{3},
J. Cortina\inst{3,2},
T. Deckers\inst{5},
S. Denninghoff\inst{2},
V. Fonseca\inst{3},
J. Gebauer\inst{2},
J.C. Gonz\'alez\inst{3},
V. Haustein\inst{4,9},
G. Heinzelmann\inst{4},
H. Hohl\inst{6},
D. Horns\inst{4},
A. Ibarra\inst{3},
M. Kestel\inst{2},
O. Kirstein\inst{5},
H. Kornmayer\inst{2},
D. Kranich\inst{2},
H. Krawczynski\inst{1,4},
A. Lindner\inst{4},
E. Lorenz\inst{2},
N. Magnussen\inst{6},
H. Meyer\inst{6},
R. Mirzoyan\inst{2},
A. Moralejo\inst{3},
L. Padilla\inst{3},
D. Petry\inst{2,6,8},
R. Plaga\inst{2},
J. Prahl\inst{4},
G. Rauterberg\inst{5},
W. Rhode\inst{6},
A. R\"ohring\inst{4},
M. Samorski\inst{5},
D. Schmele\inst{4},
F. Schr\"oder\inst{6},
W. Stamm\inst{5},
B. Wiebel-Sooth\inst{6},
M. Willmer\inst{5},
W. Wittek\inst{2}}

\offprints{plaga@mppmu.mpg.de}

\institute{Max-Planck-Institut f\"ur Kernphysik,
Postfach 103980, D-69029 Heidelberg, Germany \and
Max-Planck-Institut f\"ur Physik, F\"ohringer Ring
6, D-80805 M\"unchen, Germany \and
Universidad Complutense, Facultad de Ciencias
F\'{\i}sicas, Ciudad Universitaria, E-28040 Madrid, Spain \and
Universit\"at Hamburg, II. Institut f\"ur
Experimentalphysik, Luruper Chaussee 149,
D-22761 Hamburg, Germany \and
Universit\"at Kiel, Institut f\"ur Experimentelle und
Angewandte Physik,
Leibnizstr. 15, D-24118 Kiel, Germany \and
Universit\"at Wuppertal, Fachbereich Physik,
Gau{\ss}str.20, D-42097 Wuppertal, Germany 
%\and
%Yerevan Physics Institute, Alikhanian Br. 2, 375036 Yerevan, Armenia
%\and
%Former collaboration member
\and
Now at Forschungszentrum Karlsruhe, P.O. Box 3640, D-76021 Karlsruhe, Germany
\and
Now at Universidad Aut\'{o}noma de Barcelona,
Institut de F\'{\i}sica d'Altes Energies, E-08193 Bellaterra, Spain
\and
Now at Line Consulting AG, Hamburg}

   \date{Received; accepted}
%\titlerunning{Energy Spectrum and Composition of Charged Cosmic
%Rays}
%\authorrunning{HEGRA-Collaboration}    
\maketitle
\markboth{HEGRA coll.: Energy Spectrum and Composition of Charged Cosmic Rays}
{}

\begin{abstract}
Measurements of the lateral distribution of Cherenkov photons
with the wide-angle atmospheric Cherenkov light detector 
array AIROBICC and of the charged particle lateral distribution
with the scintillator matrix of the HEGRA air-shower detector 
complex in air showers are reported.
They are used in conjunction to determine the energy spectrum and 
coarse chemical composition of charged cosmic rays in the energy interval 
from 0.3\,PeV to 10\,PeV.
With the atmospheric shower-front sampling technique 
these detectors measure the electromagnetic component of an
extensive air shower via 
the lateral density distribution of the shower particles and
of the Cherenkov photons.
%Crucial for this purpose
%is the use of air-shower simulations,
%which should fit well the experimental observables.
The data are compared with events generated with the CORSIKA
program package with the QGSJET hadronic-event generator.
%Within the experimental uncertainties we find a good agreement
%between data and simulations. 
Consistency checks performed with primary energy-reconstruction
methods based on different shower observables indicate
satisfactory agreement between these extensive air shower simulations and
the experimental data. 
This permits to derive results concerning the energy spectrum and
composition of charged cosmic rays.
\\
The energy spectrum features a so called ``knee'' at an energy
of $E_{\rm knee}$=${\rm 3.98^{+4.66}_{-0.83} (stat) \pm 0.53 (syst)}$\,PeV. 
%a The integral flux of CR with energies larger than the knee energy is
%a found to be  
%a ${( \rm 2.5 \pm 0.5 (stat) ) \cdot 10^{-7}\,(m^2\,s\,srad)^{-1}}$.
Power law fits to the differential energy spectrum yield 
indices of ${\rm -2.72^{+0.02}_{-0.03} (stat)  \pm 0.07 (syst)}$ below and 
${\rm -3.22^{+0.47}_{-0.59} (stat) \pm 0.08 (syst)}$ above the knee.
\\
The best-fit elongation rate 
for the whole energy range is determined to
78.3 $\pm$ 1.0 (stat)
$\pm$ 6.2 (syst) g/cm$^2$. At the highest energies
it seems to decrease slightly.
%following numbers are 
%the mean of the four methods and the mean of the four errors
%in the plot where \lambda is changed by 1.05/0.95
The best-fit fraction of light nuclei decreases
from ${\rm 37 ^{+28}_{-21} \%}$  (combined\,
statistical\, and\, systematic) to 
${\rm 8 ^{+32}_{-8} \% }$ (combined\,
statistical\, and systematic) \ 
in the energy range discussed here. 
A detailed study of the systematic errors reveals that a non-changing
composition cannot be excluded.
         
\end{abstract} 
%\keywords{cosmic rays}
%
%________________________________________________________________

\section{Introduction}
\label{sec_intro}

The origin of extra solar cosmic rays (CR) is one of the important 
unresolved astrophysical questions. 
Galactic shell type supernova remnants (SNR) have been proposed as plausible
acceleration sites for cosmic rays up to  energies of several PeV
(\cite{theo}) and - for very massive
SN progenitors - to even higher energies (\cite{biermannprd}).
Recently direct experimental evidence for electron acceleration 
in these objects has been found in the X-ray 
(\cite{eacx1}; \cite{eacx2}; \cite{eacx3}) and TeV $\gamma$-ray range
(\cite{eact}).
Somewhat surprisingly, similar searches for evidence of hadron 
acceleration have only yielded upper limits on the expected $\gamma$-ray
emission from the interaction of the hadrons with interstellar matter
up to now 
(\cite{hac1}; \cite{hac2}; \cite{hac3}). \\
An indirect approach to distinguish between different theoretical 
models aiming to describe the acceleration of charged cosmic rays (CR)
is to measure the energy spectrum and composition of CR 
and compare the results with model predictions.
Here the energy regime around the so called ``knee'' between 1 and
10\,PeV is especially interesting(\cite{watson}). 
In this energy range the all-particle CR energy spectral slope
- that is constant within measurement errors for lower energies -
suddenly increases.
The riddle of the origin of the knee and of the cosmic radiation 
with energies exceeding it, is not yet finally resolved. 
The following general solutions have been discussed:
\\
1. The change in index is due to some propagation 
effect in an ``original'' cosmic-ray population that
displays an unbroken power law from low energies 
up to energies above the knee. The most 
popular idea is that the energy dependence
of the diffusion constant of cosmic rays in the Galaxy
could change in the knee region (\cite{peters}; \cite{ptuskin}).
Because of the dependence of the
diffusion constant on the nuclear
charge Z a modest decrease in the fraction of ``light''
elements (hydrogen and helium)
would be expected. In the simplified chemical model we
use below (heavy elements
modelled by 65 $\%$ oxygen and 35 $\%$ iron, light elements
by 40 $\%$ hydrogen and 60 $\%$ helium)
the fraction of light elements would be expected
to decrease from an assumed value of 60 \%
below the knee to 43 \% above the knee.
In such a model (barring a special
cancellation of effects) the knee is expected 
to be a relatively smooth
feature, extending over about a decade in energy.
A principal problem with this approach is
that no plausible Galactic source of cosmic rays has been identified
which is quantitatively
capable of producing the ``original'' cosmic-ray population.
\\
2. The knee signals in some way the maximum energy
for the sources responsible for low energy cosmic rays.
The cosmic rays at higher energies could be ``re-accelerated''
low energy cosmic-rays, e.g. at the shock front of a Galactic
wind (Jokipii $\&$ Morfill 1987) or an ensemble of shock fronts in
clusters of massive stars (\cite{bykov}). 
In this case a phenomenology similar
to the diffusion model in the previous paragraph would be expected.
Alternatively, above the knee a completely new population
of cosmic rays dominates. In this case typically dramatic
changes in chemical composition are expected, e.g. to pure hydrogen
in the extragalactical model of Protheroe (\cite{protheroe})
and nearly pure heavy elements (fraction of light elements
$<$ 0.3 far above the knee) in a model with special SNRs by
Stanev et al. (\cite{stanev}). The special properties
of these new sources could in principle allow to understand a 
knee relatively ``sharp'' in energy.
\\
To definitely discriminate between
a composition changing
as expected in models with an
energy dependent diffusion constant (discussed
above under 1.) and an unchanging composition,
it is necessary to achieve an
error of smaller than $\pm$ 10 $\%$ in the experimental
determination of the fraction
of light elements in the total cosmic radiation above the knee.
\\
While the cosmic-ray composition and energy spectrum are well known 
from direct balloon and space-borne observations 
up to energies of 
about 100\,TeV, no general agreement has been reached at higher energies
(\cite{watson}).
The results obtained for CR around the ``knee'' suffer seriously
from the fact that due to the low flux of CR above 1\,PeV, only large
ground based installations observing the extensive air showers (EAS)
induced by cosmic rays in the atmosphere can provide experimental 
data.
However the sensitivity of EAS observables to the mass of the 
primary CR is weak. The analyses are rendered even
more difficult due to theoretical uncertainties concerning the 
high energy interactions in the atmosphere 
(\cite{mccomp}; \cite{gaisser}).\\
Here we present an analysis of EAS between 300\,TeV and 10\,PeV
which restricts to observables related to the electromagnetic shower 
component.
%The analysis methods were developed and the experimental data
%compared to simulations obtained with the QGSJET event
%generators within the CORSIKA program package (\cite{cors1}; 
%\cite{cors2})).
In the following sections the experimental setup (section \ref{sec_exps}),
the Monte-Carlo simulations (section \ref{sec_mc}),
the event reconstruction 
(section \ref{sec_reco}) and analysis methods (section \ref{sec_anal})
are described.
Sections\,6 presents the results concerning the CR energy spectrum  
and the coarse mass composition.
A more detailed study of the
systematic errors and a discussion of methods to analyse the composition
without relying on the absolute penetration depth are
discussed in section 7.
The paper ends with conclusions in section\,8. 

%__________________________________________________________________

\section{The experimental Setup}
\label{sec_exps}

The air-shower detector complex HEGRA covers an area of 
180$\cdot$180\,m$^2$ at a height of 2200\,m a.s.l.\ 
(790\,g/cm$^2$) (\cite{hegra};  \cite{air};
\cite{kraw96} ; \cite{rhode}).
In the present analysis only data of the scintillator array 
and part of the AIROBICC array were used.
The former consists of 243 huts with plastic scintillators
of an area of 0.96 m$^2$, 
covered with 5\,mm of lead on a grid
with 15\,m spacing (with a denser part 
with 10 m spacing in the centre of the array).
The part of the latter used in this analysis is formed by 49 open 
photo-multipliers fitted with Winston cones, restricting the 
viewed solid angle to 0.835 sr and 
measuring the air Cherenkov light of EAS on a grid 
with 30\,m spacing.
The gain nonlinearity of all components 
in the Cherenkov-light measurement was carefully checked,
both with a LED light source with variable light intensity
and a direct charge source. While the used Cherenkov light
photomultiplier tubes
were found to be linear, the used amplifier showed an antilinearity
(gain rises with input amplitude)
which was corrected in the data analysis.
Above 10 PeV the amplifier begin to show signs of saturation
and therefore no data above 10 PeV are included in the present
analysis.
The trigger conditions used for the data analysed here demand 
a signal from at least 14 scintillator or 6 AIROBICC stations
within 150\,ns.
This corresponds to an energy threshold for 
primary protons and iron nuclei of 
25 and 80\,TeV respectively. \\ 
%anThe absolute amount of Cherenkov light was calibrated by 
%ancomparing the result of the energy reconstruction
%anusing AIROBICC data only with the primary energy derived from the 
%anscintillator and AIROBICC data on an event by event basis.
%anMore details are given in section\,\ref{erecmethods}.

\section{Monte Carlo Simulations}
\label{sec_mc}

EAS events were simulated using the CORSIKA code in its versions
5.20 with the QGSJET/GHEISHA options (\cite{cors1}; 
\cite{cors2})).
This generator is based on the quark-gluon string model (QGS)
with an allowance for semihard processes (JET) (\cite{kalmykov}). 
Complex nuclei were treated with the 
``complete fragmentation'' ansatz.
The energy cutoff for particles of the electromagnetic cascade
was set to 3\,MeV.
Proton, helium, oxygen and iron induced showers were produced 
at zenith angles of 0$^{\circ}$,6$^{\circ}$,
12$^{\circ}$ and 18$^{\circ}$
at discrete energies between 300\,TeV and 10\,PeV
(4400 independent showers of which 1000 are above 1 PeV) 
as well as an independent sample with a continuous
energy distribution at zenith angles of 6$^{\circ}$ and 12$^{\circ}$
between 50 TeV and 1\,PeV (4330 independent 
showers) and following a power law
of E$^{-1}$ between 2.5 PeV and 6.5 PeV (240 showers).
The events continuously distributed in energy were 
spectrally weighted and used in the fits 
to infer the chemical composition (section \ref{chempos}),
while the samples with discrete energies were employed
to develop the reconstruction methods and to correct
the results obtained with biased estimators of the primary energy
(section \ref{erecmethods}).
%aIn total the MC sample contains roughly 11000 independent air showers
%a(about 7500 above 300\,TeV including 750 showers
%aabove 1\,PeV). 
Note that the simulation of an EAS induced by a 1\,PeV primary proton
including the Cherenkov light production requires about 3\,h
CPU time on a 300\,MHz Pentium-II PC (during the same time the HEGRA
experiment registers more than 350 showers with energies larger 
than 1\,PeV). 
\\
The detector performance was modelled with two independent detector 
simulations:
a full detector simulation (\cite{martinez}),
and an empirical simulation using measured 
response functions ( \cite{haus}; \cite{hhsimul}).
Each independent generated EAS was used 20 times with core positions
inside and outside the HEGRA area to take into account 
the detector related fluctuations of observables and 
to check the event selection criteria. With the standard cuts described
below, each shower was used on the average two times in the 
discrete and once in the continuous MC sample.
\\
Special care has been taken to simulate the 
density profile and absorption features of the atmosphere 
above La\,Palma correctly.
Weather balloon measurements as well as comparisons between 
TeV photon data registered by the HEGRA system of
imaging air Cherenkov telescopes and simulations were employed 
for this purpose (\cite{conrad};\cite{cts}). The shower development and
light emission were modelled with the U.S. standard atmosphere,
and the light propagation was then simulated with a special
program ELBA (\cite{haus}), assuming a tropical maritime atmosphere 
for the summer.
This atmosphere is a good approximation for the conditions
at Tenerife,
an island neighbouring the experimental site (\cite{conrad}).

\section{Event Reconstruction and Data Selection} 
\label{sec_reco}

The core position of an EAS is reconstructed independently from 
the data of the scintillator matrix and from AIROBICC where the latter 
data allow to tag core positions beyond the HEGRA boundary.
If the core position lies inside the area covered with detector elements
the scintillator derived core coordinates have a resolution
of ${\rm \sigma(core) = 2 (5)\,m}$ for
protons (iron) at energies above 300\,TeV
(a little  more accurate compared to AIROBICC mainly due to the
smaller grid distances of the scintillator huts).
\begin{figure}[ht]
\vspace{0cm}
\hspace{0cm}\epsfxsize=8.8cm \epsfbox{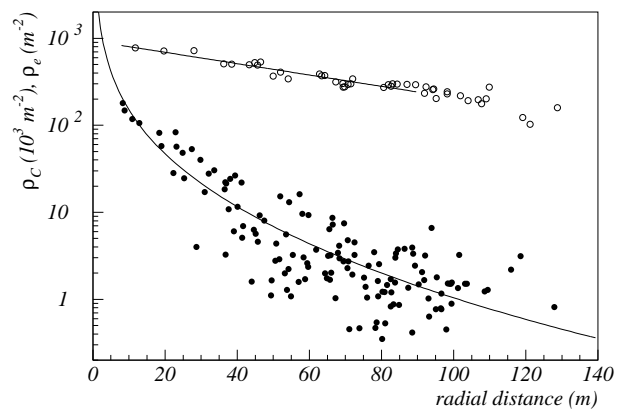}
\vspace{0cm}
\caption{
The Cherenkov-light
($\rho_C$, open circles, in 10$^{3} \times$ photons/m$^2$) and 
the actually measured (under lead coverage, see text)
charged-particle
density ($\rho_e$, full circles, in 1/m$^2$) as determined in
a single observed shower. Each open (closed) circle
corresponds to the light (charged particle) density
determined by one AIROBICC (scintillator) station.
The NKG function has been fitted to the charged-particle
distribution
and an exponential function
to the Cherenkov light distribution
in a 7.5 - 90 m radial interval (full lines). 
The energy
of this event was 2.1 (2.8) PeV if induced by a 
proton (iron) nucleus.           
}
\label{showerpic}
\end{figure} 
The direction of the primary particle is reconstructed nearly 
independently from the scintillator and AIROBICC arrival time 
data (where the scintillator derived core position is used here).  
\\
The particle density measured by the scintillator array is fitted
by the NKG formula (\cite{nkg}) with a Moli\'ere radius
of 106 m, yielding the shower size $N_s$ and an
{\em age} value.
$N_s$ is a factor $\approx$1.6 larger than $N_e$ (the ``true''
shower size at detector level, denoting the number of charged particles
above a kinetic energy of 3\,MeV) due to 
coverage of the detector huts with a lead layer
and the fact that the NKG function
does not correctly parametrise the electromagnetic part of hadronic showers.
\\  
%The exact value depends on the distance between detector and shower 
%maximum.
The dependence of the Cherenkov photon density ${\rm \rho_c}$, as measured
by AIROBICC, on the distance r from the shower axis can be
well described by an exponential in the region
20\,m $<\,$r\,$<$\,100\,m
(\cite{hillas}):
\begin{equation}
{\rm \rho_c(r)\,=\,a \cdot exp(r\cdot {\em slope})}.
\end{equation} 
The parameter {\em slope} (in units of [1/m]) is the most 
important one in our analysis methods.
As an illustration Fig.\ref{showerpic} shows the lateral
charged-particle and Cherenkov-light distributions for a single shower.
\\ 
The amplitude calibration of the scintillator array is done 
for samples of 50000 events by 
comparing the ADC spectra of the individual huts
- which display a single peak essentially corresponding
to the energy deposited by minimum ionising electrons 
and muons -
with the result of
MC simulations for identical conditions.
The absolute amount of the air Cherenkov light registered by 
AIROBICC was calibrated by comparing the energy inferred
from the
lateral Cherenkov light density in the spectral range from 300 nm to 500 nm
registered at a shower core distance of 90\,m 
(referred to as $L_{90}$ in the following)
and the energy derived from $N_s$ and 
{\em slope} in the interval 
${\rm 2.5 < log_{10}(E(N_s, } slope {\rm )/TeV) < 2.75}$
(refer to section \ref{erecmethods} below for energy reconstruction methods).  
The absolute Cherenkov-light 
calibration thus depends on the CR mass composition,
because we do not apply a primary-mass independent energy reconstruction
here.
%anWe used our measured composition for the final calibration.
We used the low-energy composition at 100 TeV as specified by
Wiebel-Sooth et al.(\cite{wiebel}) (60 $\%$ light elements, see below
section \ref{chempos} for details) for this calibration. 
If a pure proton (iron) composition is assumed 
the energy reconstructed from Cherenkov light alone is shifted by 
3 (13)\% to higher (lower) energies. 
\\ 
To select well-reconstructed events the EAS core positions
and directions as reconstructed with AIROBICC and the scintillator 
matrix are demanded to be consistent. 
Additional cuts ensure the quality of the directional as well as the 
fits to the lateral particle and Cherenkov light density 
distributions. 
Events with the true shower-core position  
within the HEGRA array boundaries for the detector
components used in this analysis (distance to edge of array $>$ 10 m) 
and a zenith angle below 15$^0$
are used for the further analysis. 
The efficiency to select EAS events with true core positions in the 
regarded 160$\cdot$160\,m$^2$ area is about 98\% for 
primary energies above 300\,TeV (independent of the primary mass).
The contamination of the sample with EAS, where the true shower cores
lay beyond the HEGRA boundary but which were erroneously reconstructed 
to fulfil the cuts is less than 1\% from 
our simulations.
\\
Nights with perfect weather conditions are selected by 
data of the Carlsberg Meridian Cycle (\cite{carls}) and by comparing the 
Cherenkov light measurements with data from the scintillators
for samples of 50000 events (accumulated in about one hour with the 
used setup and trigger conditions).
The data set solely contains nights without any technical problems 
of the used detector stations.
In total it comprises (dead-time corrected) an on-time of 
208\,h. This corresponds to about 150000 events 
after all cuts with an energy above about 
100\,TeV and a zenith angle below 15$^0$.

\section{The Analysis Methods}
\label{sec_anal}

Monte Carlo simulations revealed that the distance 
${\rm d_{\rm max}}$ between detector and 
shower maximum
(defined as the point in the shower development with the
maximal number of charged particles)
can be reconstructed independently of the primary mass 
with the shape parameter {\em slope} of the lateral Cherenkov light 
density distribution: 
\begin{equation}
{\rm d_{\rm max}\,=\, } [{\rm 680\,+\,}slope 
{\rm \cdot\  20880\,m}] {\rm \,g/cm^2}.
\label{dimax}
\end{equation}
From this relation the distance to the shower maximum
is determined with a resolution 
(i.e. root mean square
(RMS) of a (d$_{\rm max}$(true)-d$_{max}$(reconstructed) distribution)
ranging from 40\,g/cm$^2$ at 300\,TeV
to 20\,g/cm$^2$ at 10\,PeV (including all detector effects but no
systematic error in the mean d$_{\rm max}$). 
The most important technical improvement in the data presented
here to previous experiments is that
these values are distinctly smaller than the width
of natural shower fluctuations of proton
induced showers in the atmosphere (see below Table \ref{table1}).
This makes the shape of the penetration-depth
distribution a sensitive parameter for the chemical composition (see
section \ref{sec_reli}).   
Simple geometrical
relations permit to
infer $X_{\rm max}$, the depth of the shower maximum in the 
atmosphere, from $d_{\rm max}$.
Relation \ref{dimax} is only weakly energy dependent (\cite{axel}).
This dependence is neglected here.
\subsection{Energy Reconstruction}
\label{erecmethods}
\begin{figure}[ht]
\vspace{0cm}
\hspace{0cm}\epsfxsize=8.8cm \epsfbox{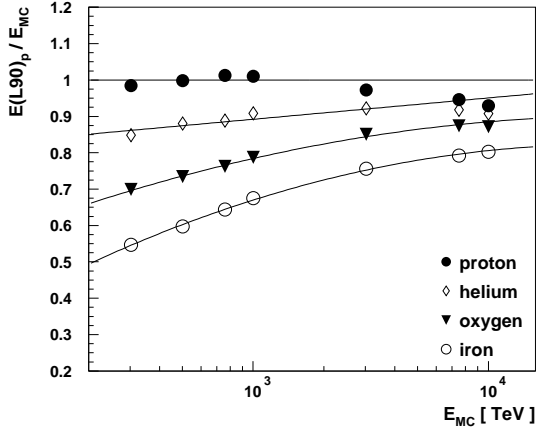}
\vspace{0cm}
\caption{
The bias of the energy reconstruction as a function of
primary energy for different primary masses.
Shown is the ratio of the reconstructed energy with 
method 3, divided by the true energy (from the Monte-Carlo
simulation).
Very similar results are obtained from 
$N_s$ and {\em slope}.
The lines show fits used for convolution procedures
to determine the final results (see text).
}
\label{ereco}
\end{figure}

\begin{figure}[ht]
\vspace{0cm}
\hspace{0cm}\epsfxsize=8.8cm \epsfbox{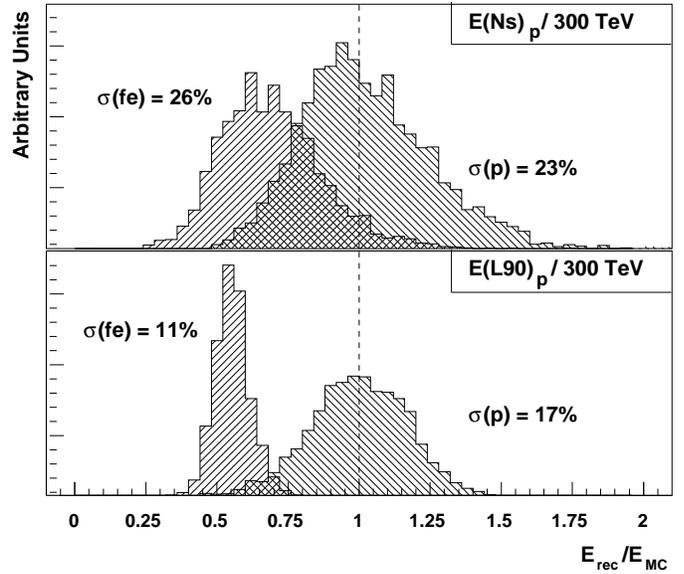}
\vspace{0cm}
\caption{
Distribution of the ratio of reconstructed to MC generated 
energy (300 TeV) for two primaries (left distribution: iron nuclei,
right distribution: protons)
and the two different energy 
reconstruction methods discussed in the text.
Each distribution is normalised to the same area. 
}
\label{rms.single}
\end{figure}

\begin{figure}[ht]
\vspace{0cm}
\hspace{0cm}\epsfxsize=8.8cm \epsfbox{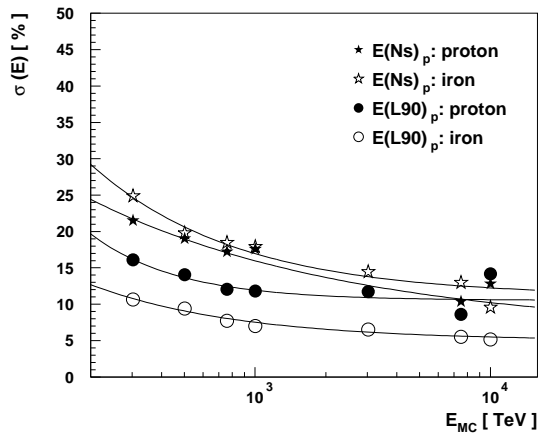}
\vspace{0cm}
\caption{
The energy resolution obtained for different primary nuclei 
as a function of the generated MC energy.
$N_{\rm s}$ in brackets denotes an energy reconstruction that combines 
the measured shower size at detector level and {\em slope},
$L_{\rm 90}$ denotes the results obtained from the Cherenkov light
density alone. The
energy reconstruction was always 
performed assuming that the primaries
are protons (referred to as methods 1 (stars in the Figure) and 3 
(dots in the Figure) in the text; this fact is symbolised
by the subscript ``p'' on the energy). The ``proton'' (full symbols)
``iron'' (open symbols)
after the colon indicate the primary for which the energy
resolution was determined.
The lines show fits used for convolution procedures
to determine the final results (see text).
}
\label{erms}
\end{figure}

Methods have been developed to reconstruct
the primary CR energy from the scintillator 
and AIROBICC data {\it independently} of the primary mass with an 
accuracy better than 35\%
(\cite{axel}; \cite{juan}; \cite{juanthesis}).
However, these methods lead to a 
relatively strong correlation between reconstructed energy
and $X_{\rm max}$ 
(showers with a maximum position that fluctuated to smaller
values compared to the mean $X_{\rm max}$, are reconstructed
with higher energies).
In order
to infer the chemical composition, a careful modelation of the response
function between the variables $N_s$ (or $L_{90}$) and {\it slope} 
on the one hand and energy and composition (or penetration depth)
on the other hand ( e.g. via two-dimensional regularised unfolding)
is then necessary. Such procedures have been employed
in some analyses of HEGRA data (\cite{barbara}; \cite{harald}).
Two reasons lead us to
prefer to circumvent the mentioned
problem with the use of two simpler energy estimators here.
%These estimators are based on $N_{\rm s}$ and {\em slope}
%and $L_{\rm 90}$ respectively. Both are used under the assumptions
%that all primary CR are protons and iron nuclei. These
%extreme assumptions lead to a bias which is then corrected for.
\\
One reason is that the methods described below are based on 
physically transparent properties of air-showers
inferred from the Monte-Carlo simulations.
Whether these properties really hold, is 
tested to some degree using different energy estimators
with different biases and comparing the obtained results.
These consistency checks are an important 
advantage over more refined and
complete methods when it is doubtful how well the Monte-Carlo 
simulation describes the data.
The other reason is
that the Monte Carlo statistics at the highest energy
is still rather limited and mean shower properties
are inferred with higher certainty than a complete
response matrix.
The mass independent energy reconstruction methods will be 
applied to the data in a forthcoming publication 
together with a discussion of the influence of different EAS
simulations.
The energy estimators used in this paper and described 
below are based on $N_{\rm s}$ and {\em slope}, or $L_{\rm 90}$. Both
estimators are used under the assumption that all primary
CR are either protons or iron nuclei. These extreme assumptions
lead to a bias which then has to be corrected for.
\\
Using $N_{\rm s}$ and {\em slope}
the energy of the primary cosmic-ray nucleus is reconstructed 
in two basic steps here (\cite{axel98}; \cite{juan}; \cite{rainer95}): 
first {\em slope} (a measure of the distance to the shower maximum)
is combined with $N_{\rm s}$ to estimate the number of 
particles in the shower maximum which is proportional to the 
energy contained in the electromagnetic component of the EAS.
In the second step a specific primary mass is assumed;
with the assumption of primary proton (iron) we denote
the methods as 1 (2).
This allows to calculate the primary energy from the energy 
deposited in the electromagnetic component.
The following relation was used in our analysis:
\begin{equation}
log(E[TeV]) = a \cdot log(N_s({\rm max})) + b.
\end{equation}
Here $a,b$ were obtained from 
the discrete Monte Carlo data as 0.965,$-$2.545 (0.890,$-$2.010)
for protons (iron).
{$N_{\rm s}({\rm max})$} 
is the shower size at the maximum of shower development
and is inferred from $N_s$ as:
\begin{equation}
N_{\rm s}({\rm max})/N_{\rm s} = a_0 + a_1 {\em slope} + a_2 {\em slope}^2 +
a_3 {\em slope}^3
\end{equation}
with $a_n$ given as (0.57833,-85.146,6181.8,-71054) 
for all primary nuclei.
This procedure is valid because the shape of the shower
development is only weakly dependent on the mass of the
CR nucleus A, especially after the shower maximum (\cite{axel}).
Only the fraction of the total energy fed
into the electromagnetic cascade depends on A for
a given energy per nucleus. The comparison of the results assuming 
initially proton and iron primaries is a consistency
check for the dependence of shower size at the maximum of
shower development on the energy per nucleon. 
\\
Alternatively the energy is reconstructed from the 
AIROBICC data alone (method 3 (4) with the assumption
of proton (iron) primaries).
Here it turns out that $L_{90}$ is a good 
estimator of the energy contained in the electromagnetic EAS 
cascades.
From simulations the relation 
\begin{equation}
log(E[{\rm TeV}]) = a \cdot log(L_{\rm 90}[{\rm photons/m^2}]) + b
\end{equation}
was derived,
where the coefficients $a$ and $b$ are given
as 0.958,$-$1.810 (0.840,$-$1.061)
for primary protons (iron).  
{\em slope} 
(the parameter used to estimate the primary mass 
composition, see next section) is not involved in this energy reconstruction.
For ease of reference the four energy-reconstruction methods
are summarised in Table \ref{taberec}.
\begin{table}[h]
\vspace{-10pt}
\caption{Summary of the energy reconstruction methods 1-4 as
discussed in the text.}
\label{taberec}
\begin{center}
\begin{tabular}{lccc}
\hline\hline
Method \# & parameter used & primaries assumed to be \\
\hline
 1 & $N_{\rm s}$ and {\em slope} & protons  \\
 2 &   $N_{\rm s}$ and {\em slope} & iron nuclei   \\
 3 &  light density $L_{\rm 90}$  & protons  \\
 4 &   light density $L_{\rm 90}$ & iron nuclei  \\
\hline
\hline
\end{tabular}
\end{center}
\end{table}
The agreement of analyses based on $N_s$ and $L_{90}$ 
is a consistency test for the accurate description
of the longitudinal shower development
by the Monte Carlo simulation.
\\ 
Naturally (because the fraction of the primary energy deposited in 
electromagnetic cascades depends on the energy per nucleon of the 
primary particle)
the mean of the calculated energy is only correct for the assumed 
particle type (Figure\,\ref{ereco}).
The biases shown in Figure\,\ref{ereco} have to be corrected for,
to derive the real energy spectrum and CR mass composition 
from our measurements (see section \ref{chempos}, \ref{espec}).
%a This clear disadvantage compared to a mass-independent energy reconstruction
%a has to be dealt with in order to lower systematic uncertainties   
%a related to possible imperfections of EAS simulations.
In order to check that our final results do not depend on the 
assumed primary-particle mass, we shall always compare the results 
based on the four energy reconstruction methods below.
\\
The distribution of the reconstructed energy compared to the 
simulated energy is shown for examples in Figure\,\ref{rms.single}.
Note that the energy reconstruction from $L_{90}$ alone 
shows Gaussian distributions while the energy obtained
from $N_s$ and {\em slope} exhibits tails to high values 
which have to be taken into account properly in the analyses.
Figure\,\ref{erms} shows the relative energy resolution achieved 
for different primary particles 
and energy reconstruction methods 1 and 3.
If $N_s$ is involved in the energy reconstruction 
the energy resolution is limited by the experimental accuracy
of the shower size determination at the detector level.
Due to the smaller $N_s$ of iron compared to proton induced 
EAS the accuracy of the energy reconstruction for iron showers is a 
little worse than for proton showers.
The energy resolution obtained from $L_{90}$  
is mainly determined by fluctuations in the shower development
(being larger for proton than iron induced showers) and could not
be decreased much by improving the detector.
\\
In all analyses below
we bin the data in six equidistant energy intervals
from log$_{10} E_{\rm reconstructed}$ [TeV]=2.5 to  
log$_{10} E_{\rm reconstructed}$ [TeV]=4.0 (see e.g. Fig.\ref{fitmax}).
Event samples defined to contain events in a certain
reconstructed-energy interval for the four
energy-reconstruction methods then contain 
events with different true primary energies.
It should always be kept in mind that 
these four samples
are not independent because they are all based on the
same total data sample.
\begin{figure}[ht]
\vspace{0cm}
\hspace{0cm}\epsfxsize=8.8cm \epsfbox{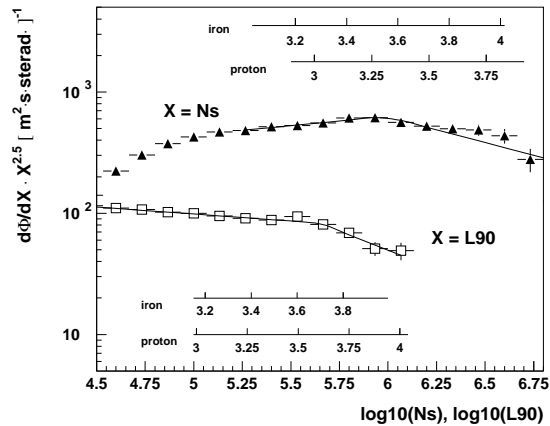}
\vspace{0cm}
\caption{
%divide Ns by conversion factor HH (ca. 1.6) before plot
The differential shower-size spectrum 
%(in units of charged particles
%above 3 MeV) 
and ``light-density at 90 m core distance (L$_{90}$)'' spectrum.
The values used for the construction of these spectra, were 
employed for the energy reconstruction. The full lines 
indicate the best fit in the range
5.3-6.8 and 4.5-6.1 for
$N_s$ and $L_{90}$ respectively. Two power laws
which meet in a single flux value
(``the knee'') were assumed. The best-fit power
law indices are (-2.35/-2.92) and (-2.61/-3.13) (before/after) the knee
at a position of log$_{10}$($N_s$)/log$_{10}$($L_{90})$ = 
5.99/5.64 for the shower size/light density respectively. 
The energy scales were derived under the assumption that
the primaries are all protons, resp. iron nuclei
for shower size (upper scales) and light density (lower
scales).
}
\label{nsl90}
\end{figure}
\subsection{Chemical Composition}
\label{chempos}
The composition of CR is determined by analysing 
the EAS penetration depth ($X{\rm _{max}}$) distributions 
in intervals of the reconstructed primary 
energy.
Information is contained
in the differences of the mean $X{\rm _{max}}$  
values for different primaries (protons penetrate about 100 - 130\,g/cm$^2$
deeper than iron in the energy range considered here) and also 
in the the different fluctuations of the shower maxima 
position. Including experimental resolution
we obtained  
RMS($X{\rm _{max}},p)\,=\,90\,g/cm^2$ and 
RMS($X{\rm _{max}},Fe)\,=\,50\,g/cm^2$ at 1\,PeV).
The RMS values of the depth distributions of
Monte-Carlo events
slightly decrease with rising energy, an effect
that is partly due to 
an improving measurement of {\em slope}.
\\
We perform an analysis which uses both of these
parameters in one fitting procedure. As the error
from such an analysis turns out to be already quite 
large, we do not perform an analysis based on mean
penetration depth alone. An analysis based mainly
on the fluctuation of penetration depths is discussed in section
\ref{sec_reli}.
\\
The present data are not sensitive enough on the chemical
composition to allow a analysis with several
mass groups; therefore 
we restrict ourselves 
to a determination of the fraction of light nuclei (protons 
and helium) by fitting the expected
to the measured $X{\rm _{max}}$ distributions.
To define the MC expectations for light nuclei, the generated 
distributions for primary protons and helium nuclei are added
with weights of 40\% and 60\% (the ratio derived from direct measurements
at energies around 100 TeV (Wiebel-Sooth et al. \cite{wiebel})).
The distribution of heavier nuclei is constructed analogously by summing
65\% oxygen and 35\% iron induced EAS.
Variations in this ratio at higher energies are
possible and are an additional potential source for
systematic errors that is not further considered below.
\\ 
The spectrally weighted Monte-Carlo data
are fitted to
the measured penetration-depth distributions
for each of the four energy-reconstruction methods.
Because spectrally weighted
Monte-Carlo data were available only for
the energy bins log$_{10}$ $(E_{\rm reconstructed})$ = 2.5 - 2.75
and 3.5 - 3.75 the energy bins between 2.5 - 3.25 (3.25 - 4)  were fitted 
with the former (latter) 
distribution. The MC events used in energy intervals other than
the two for which the simulations were done, were
shifted in the mean penetration depth
according to the elongation rate of the various elements.
To avoid any systematic uncertainties related to imperfect 
parameterisations of the MC distributions and to take into 
account the statistical uncertainty of the simulated event sample
we directly fit the MC generated distributions to the experimental data.
\\
Due to the primary
dependent energy-reconstruction 
method the results
for the
``fraction of light nuclei'' (abbreviated ``(p + $\alpha$)/all'' below)
are biased.
The results for these fits in the chosen energy bins are shown for
method 3 in Fig.\ref{fitmax}. The obtained (p + $\alpha$)/all  
ratios are then corrected for the A dependent bias which
is illustrated in Fig.\ref{ereco}. 
The correction can be
described as a single overall factor for the (p + $\alpha$)/all
ratio for each energy bin 
- rather than a transformation of the penetration depth
distribution -
to a good approximation because of the independence
of our energy reconstruction methods 
of $X_{\rm max}$ as discussed in section \ref{erecmethods}.
These correction factors were derived from spectrally weighted
Monte-Carlo data via determining the true 
(p + $\alpha$)/all in the Monte Carlo
that yields the fitted biased (p + $\alpha$)/all in the given 
reconstructed-energy bin.
In this way the ratio of biased to
true (p + $\alpha$)/all at the true mean energy
of the Monte-Carlo showers in the energy bin for a given energy
reconstruction method is obtained.
As an illustration the correction factors 
for the case of energy reconstruction
method 3 are shown in Table \ref{table2}.
\\
For the spectral weighing of the Monte-Carlo sampling a primary-spectrum 
as obtained from low-energy measurements 
(Wiebel-Sooth et al. \cite{wiebel}) with a power
law index of $\alpha$=$-$2.67 and a ``knee'' 
at 3.4 PeV with a change in the power-law
index to $\alpha$=$-$3.1 was assumed. 
An iterative repetition of this procedure with the energy
spectrum as inferred below from the present data is possible.
However,
it was found that the contribution to the systematic error introduced
by not performing the iterations is negligible for the initial
parameters chosen.   
%r The correction factor C for an energy bin
%r is then given as (E$_{\rm reconstructed}$/E$_{MC}$)$^{-\alpha}$,
%r where E$_{\rm reconstructed}$ is the mean energy of all
%r events in a bin.
\\
Two Monte-Carlo samples were used for bias corrections in
this work,
the Monte-Carlo sample 
with events continuously distributed in energy, mentioned
in section \ref{sec_mc}, and a ``toy Monte-Carlo sample''
with unlimited statistics, which was created by randomly choosing 
all measured parameters (like reconstructed energy, $X_{\rm max}$ etc.)
of a shower with a given true primary energy
from one dimensional distributions inferred from the Monte Carlo
sample with discrete energies.
It was checked that the corrections obtained with these samples 
are very similar in energy regions where the continuously
distributed Monte-Carlo data were available.  
\begin{table}[h]
\vspace{-10pt}
\caption{The fraction of light
elements (uncorrected) and correction factor 
for the A dependent
bias with energy reconstruction method 3.
The uncorrected ratio has to be multiplied by this factor 
to yield the unbiased ratio.
The energy intervals are specified
as the logarithm to the base of ten in units of TeV.}
%other table ratio of (p+$\alpha$)/all, and correction factors
%other table for energy reconstruction method 3.
\label{table2}
\begin{center}
\begin{tabular}{lccc}
\hline\hline
Reconstructed energy & p + $\alpha$/light & correction factor \\
\hline
2.5 - 2.75 &  0.502 & 0.741  \\
2.75 - 3. &  0.493 & 0.763 \\
3. - 3.25 &  0.553  & 0.813  \\
3.25 - 3.5 &  0.406 & 0.800   \\
3.5 - 3.75  &  0.272 & 0.806 \\
3.75 - 4. &  0.09  & 0.820   \\
\hline
\hline
\end{tabular}
\end{center}
\end{table}
\subsection{Energy spectra, elongation diagrams and penetration
depth fluctuations}
\label{espec}
Energy spectra obtained with the four
energy-reconstruction methods were corrected
for the A dependent bias by dividing the flux
values in bins with true and reconstructed energy in the
Monte-Carlo samples. The chemical composition as determined with
the methods in the previous section is used.
These factors were
applied to the flux in each energy bin when going from reconstructed
(Fig. \ref{eint}) to true energy (Fig. \ref{eint2}).
\\
$X_{\rm max}$ as a function of true energy is 
obtained if the mean $X_{\rm max}$ is plotted
at the mean true energy
of the events in a given reconstructed-energy bin, as   
calculated with the measured chemical composition.
This procedure leads to correct results as long
as the elongation rate of different nuclei is
identical; this is fulfilled to a good approximation
for all hadron generators.
\\
The RMS of the shower penetration depth distributions
were directly calculated from the distributions
calculated with a given energy-reconstruction method,
i.e. no procedure to remove the bias was applied.
These results were compared with RMS values
from Monte Carlo data treated in the same way. 

\subsection{Experimental Statistical
and Systematic Uncertainties}   
\label{syst}
\begin{figure}[ht]
\vspace{0cm}
\hspace{0cm}\epsfxsize=8.8cm \epsfbox{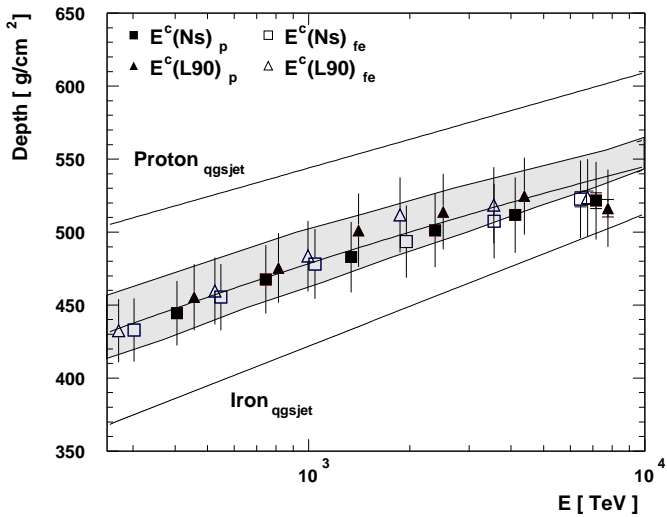}\vspace{0cm}
\caption{  
The mean shower maximum depth as a function of energy 
%a in comparison
%a to other EAS data and extrapolations from direct measurements using 
using QGSJET simulations to model the EAS development. 
To obtain an unbiased elongation plot
each data point is plotted at the true mean energy
of the events used to infer the mean $X_{\rm max}$. The
shaded region indicates the region expected from our best
fit composition within its total error. Errors are statistical
and systematical errors added in quadrature, up to the tick mark
only statistical. Up to the highest energies the systematical error
dominates.
}
\label{hmax}
\end{figure}

For the energy spectrum
the statistical uncertainties correspond to the square root
of the energy-bin contents N for the energy spectrum and the
mean $X_{\rm max}$ divided by $\sqrt{N}$ for the penetration depth.
In all other cases
statistical errors were obtained
by changing the fit parameter from its
best-fit value until the $\chi^2_{red}$ increases by 1. In case
of  best fit $\chi^2_{red}$'s in excess of 1.5 the best fit
value of the fit parameter was increased until $\chi^2_{red}$ doubled.
\\
Systematic uncertainties of the Monte-Carlo
simulation of hadronic air-showers - estimated 
by using different hadronic Monte-Carlo generators - will be 
considered in a forthcoming paper.
Here we concentrate on experimental uncertainties 
related to the {\em slope} reconstruction.
These are contributions from remaining uncertainties in the
characteristics of the 
AIROBICC amplifier (3\% uncertainty for {\em slope}) and
non-perfect knowledge of the layer structure and the light absorption   
of the atmosphere above the detector (4\% and 2\%).
Models of the atmosphere have been carefully checked using 
the large statistics of photon induced air showers which 
were registered with the HEGRA imaging air Cherenkov
telescopes in 1997 (\cite{cts}).
Added in quadrature the systematic uncertainty of {\em slope}
amounts to 5\%.
The mean $X{\rm _{max}}$ for 300\,TeV proton (iron)
induced showers is then determined with an uncertainty in the absolute
values of 
20\,(13)\,g/cm$^2$.
The uncertainties for different primaries are 
strongly correlated.
\\
For the chemical composition, the energy spectrum
and the variation of $X_{\rm max}$ with energy (elongation rate),
the systematic error was evaluated by changing all {\it slopes}
by 5 \% (the systematic error of this parameter) up or down.
The whole analysis, including energy reconstruction, was then repeated
and the deviation of the results thus obtained to the
original ones was taken as the systematic error (errors beyond
the tick mark in Figs. \ref{compos},\ref{hmax} and shaded bands in 
Figs. \ref{eint2},\ref{ediff}).
The shaded band in Fig. \ref{hmax} is obtained by
varying the best-fit composition within its total systematic and
statistical error.
In the case of the elongation rate the systematic error was found
to be dominated by the differences in the four energy-reconstruction
methods, this is dicussed in detail in the Results section
\ref{sec_elon}.
In case of the RMS of the penetration depths, the spectral
fit parameters (knee position, power-law indices) 
and the elongation rate, the systematic error
was estimated as the sample standard deviation of the best
fit parameters obtained with the four energy-reconstruction methods.
%__________________________________________________________________

\section{Results}
\label{sec_res}
In this section the methods explained
in section \ref{sec_anal} are applied to the data set
discussed in section \ref{sec_reco}.
\subsection{$N_{\rm s}$ and $L_{\rm 90}$ spectra}
Fig. \ref{nsl90} shows the $N_{\rm s}$ and
$L_{\rm 90}$ spectra. These spectra display a relatively sharp
knee at values consistent with a primary energy for the
knee as determined below.
\subsection{Energy Spectra}
\label{sec_ener}
\begin{figure}[ht]
\vspace{0cm}
\hspace{0cm}\epsfxsize=8.8cm \epsfbox{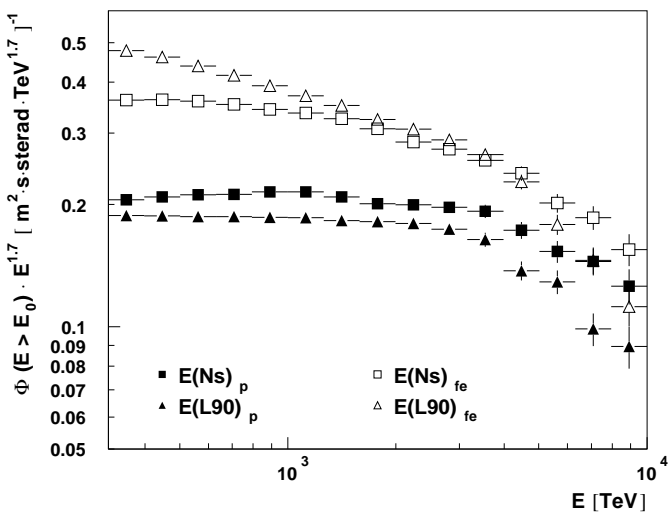}
\vspace{0cm}
\caption{
The integral energy spectra obtained with the four
energy reconstruction methods denoted with different
symbols.
Full square: Method1, open square: method 2, full triangle:
method 3, open triangle method 4. 
The biases of the energy reconstruction have not been corrected for.
}
\label{eint}
\end{figure}

\begin{figure}[ht]
\vspace{0cm}
\hspace{0cm}\epsfxsize=8.8cm \epsfbox{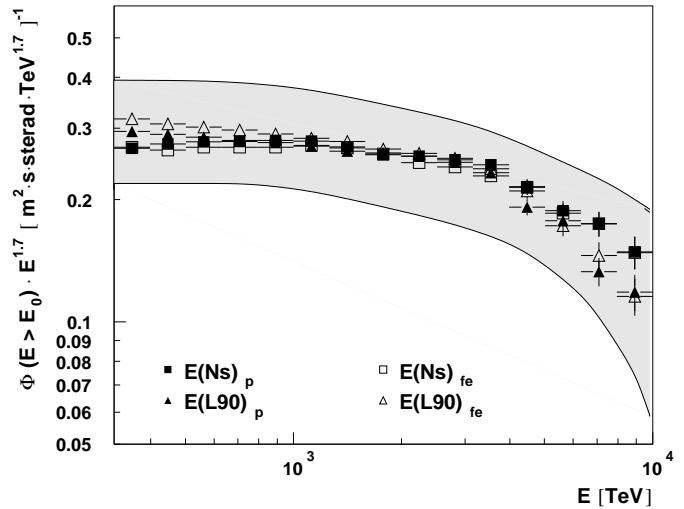}
\vspace{0cm}
\caption{
Integral cosmic-ray spectrum corrected for the A dependent bias.
The shaded area denotes the systematic error. Symbols are the
same as in the previous Figure.
%a The lines show fits to the data.
%a The shaded areas show the results which are obtained if
%a either a pure iron (top) or pure proton (bottom) 
%a composition. 
}
\label{eint2}
\end{figure}

\begin{figure}[ht]
\vspace{0cm}
\hspace{0cm}\epsfxsize=8.8cm \epsfbox{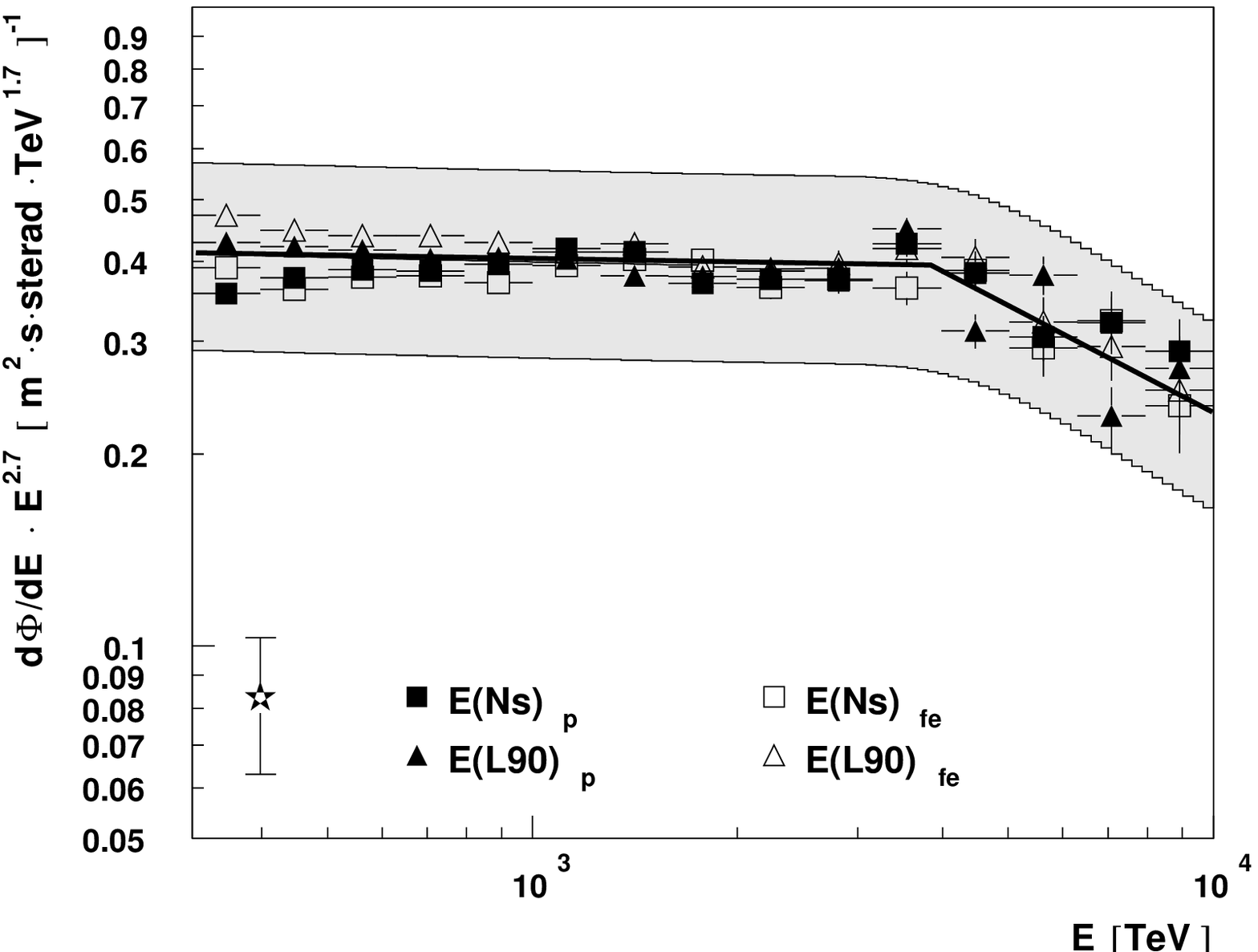}
\vspace{0cm}
\caption{  
The differential CR energy spectrum obtained
with four energy reconstruction methods.
Symbols are the same as in Fig. \ref{eint}.
The light shaded region represents systematical uncertainty.
%The HEGRA data are compared to recent measurements of the DICE
%(open circles, \cite{dice})
%and As-$\gamma$ (full dots, \cite{tibet}) collaborations.
%Extrapolations \cite{wiebel} from direct balloon data are shown as the 
%hatched area.
The ``star'' with vertical error bars shows the uncertainty of 
the HEGRA data originating from the 10$\%$ systematic 
uncertainty of the 
absolute energy scale that we estimate from the uncertainty in the 
determination of the absolute
$N_s$ scale. The full line is the best fit as described in the text.
} 
\label{ediff}
\end{figure}

Figure\,\ref{eint} displays the integral energy 
spectra uncorrected for an A
dependent bias obtained with the four reconstruction methods.
The differences in absolute normalisation
and spectral slope originate from the different 
mass dependent biases.
After the correction of the chemical bias the integral spectra  
are similar (Figure\,\ref{eint2}). This nontrivial fact
is in favour of the internal consistency of data analysed here; a
longitudinal shower development different from the one predicted by the
Monte Carlo or errors in the calibration of $L_{90}$
and $N_s$ could have spoiled the agreement of spectra
obtained with different energy reconstructions.   
 The differential energy spectrum is shown in 
Figure\,\ref{ediff}.
A steepening of the energy spectrum is visible 
around an energy of 4\,PeV.
There seems to be no ``fine structure''
in the energy spectrum around the knee in excess
of 20 $\%$. Apparent
structure with smaller amplitudes that appears in the
spectrum reconstructed with a given energy-reconstruction
method is not reproduced 
with other methods. 
This is expected due to the A dependent
bias of our energy-reconstruction methods 
(see Fig.\ref{ereco}).
Note that with these methods, a potential structure
in the energy spectrum consisting of different nuclei
is smeared out.
If two different power laws, smoothly connected
at the knee (corresponding to a ``sharp'' knee),
%a by 
%a${\rm exp(a+b\cdot E + c \cdot E^2)}$
%a over an energy interval of ${\rm \pm \Delta log_{10}(E/E(Knee))=0.25}$
are fitted to the differential spectra we obtain:
\begin{itemize}
\item a ``knee'' position 
of $E$(Knee)\,=\,${\rm 3.98^{+4.66}_{-0.83} (stat) \pm 0.53 (syst)}$\,PeV,
\vskip 0.2in
\item a spectral index of 
${\rm -2.72^{+0.02}_{-0.03} (stat)  \pm 0.07 (syst)}$ below 
and ${\rm -3.22^{+0.47}_{-0.59} (stat) \pm 0.08 (syst) }$ above the knee.
% and
%\item an integral CR flux of  
%${\rm F_I(E>E(Knee)) = (2.5 \pm 1)\cdot 10^{-7}\ 
%(m^2\,s\,srad)^{-1}}$
%above the knee energy.
\end{itemize}
%a${\rm F_I(E>E(Knee))}$ does not depend on any 
%auncertainties on the absolute energy scale.
%aTherefore this quantity is the most robust one, which
%acould be compared between different experiments.
%r It is difficult to quantitatively measure of the ``sharpness''
%r of the knee. 
The reduced $\chi^2$ values of the fits to the differential
spectrum (12 d.o.f.) were 6.75, 4.03, 3.53 and 1.47 with energy reconstruction
methods 1 to 4 respectively. Some of these values are much larger than one.
It is then difficult to specify a statistical
error; we specify the statistical errors for method 4 that has
a marginally acceptable reduced $\chi^2$ value.
The large $\chi^2$ values for the analysis with
energy-reconstruction methods
1-3 can be interpreted as an argument in favour of a knee
not absolutely sharp in energy.
However, the fact that one of the fits
is acceptable on the 90 $\%$ confidence level means
that we cannot reject the hypothesis of a ``sharp''
knee (two power laws with no transition region) within
our systematic errors.
The large statistical error on the knee position 
further indicates that
we cannot reject the hypothesis of a spectrum
without a knee in the limited energy range of this
analysis with high significance.
\\
The spectral index for the spectrum below the knee is consistent
with direct measurements at lower energies
(Wiebel-Sooth et al. \cite{wiebel})
and a recent Cherenkov-light based determination
of the spectral
index in the TeV range (\cite{hdspec}); there is therefore 
no evidence for any change in spectral index from the TeV range
right up to the knee.  
%r We only observe here that the spectrum
%r seems to be well described by a single power law
%r from 300 to 3000 TeV to within about 20 \% in absolute
%r flux. 
%r This might  
%r for the Peters model in which the light component should have
%r a knee, about a factor 5 below the mean knee position,
%r leading to a continuous steepening in this energy range.
\\

%
%a INTERPRETATION: SCHARFES KNIE ODER EHER WIE TIBET,
%a VERGLEICH MIT DIREKTEN MESSUNGEN (10% 
%a SYSTEMATIK AUFGRUND ABSOLUTER ENERGIESKALA AUS VERGLEICHEN 
%a VERSCHIEDENER MODELLE),
%a STRUKTUR IM SPEKTRUM VOR DEM KNIE?

%__________________________________________________________________

\subsection{Composition of CR}
\label{sec_comp}
\begin{figure*}[ht]
\vspace{0cm}
\hspace{0cm}\epsfxsize=16cm \epsfbox{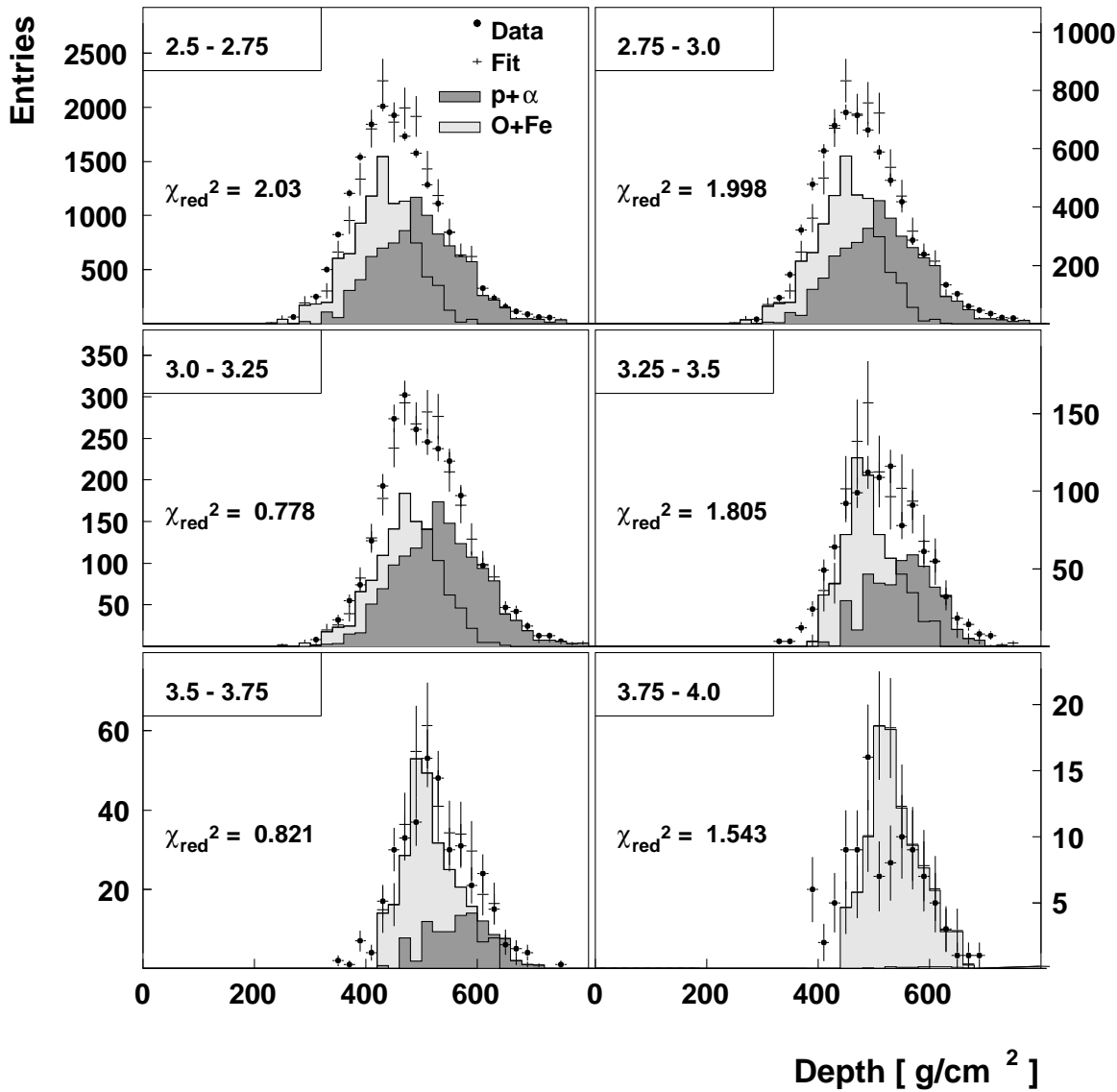}
\vspace{0cm}
\caption{  
The fit of MC expectations for light and heavier nuclei
to the measured shower maximum depth distribution in the analysed
reconstructed-energy intervals for method 3.
The numbers in the upper left corner are the 
logarithms to the base of ten of the energy-bin boundaries in TeV. 
The full dots mark the experimental data with statistical errors,
the crosses with error bars are the fitted MC distribution
(where the errors correspond to the MC statistics).
The two components fitted to the data are shown 
as dark shaded (large penetrations depths,
light nuclei) and light shaded (heavy 
nuclei) histograms.
Details of the procedure are described in the text.
}
\label{fitmax}
\end{figure*} 
The fraction of light nuclei as a function of 
reconstructed energy - obtained from the fits to the
measured penetration-depth distributions (see section \ref{chempos})-
is presented in 
Figure\,\ref{compos}.
At energies below the knee the composition is mixed and
consistent with direct measurements around 100 TeV, namely
($p + \alpha$)/all = 0.54 $\pm$ 0.08 (\cite{watson}).
The data points seem to indicate a gradual enrichment
of heavy elements above about 1 PeV though
the error bars are large (remember that there
are only six {\it independent} data points).
We will discuss in section \ref{sec_reli} how
reliable the qualitative conclusion of a gradual enrichment
in heavy elements is within our systematic errors.
The data rule out a predominantly light composition at
all energies and does not give evidence for a drastic change
of composition at the knee.

%a INTERPRETATION (GROSSE CHI2 WEGEN KOMBINATION ALLER MCS OBERHALB
%a VON 500 TEV?).
%a The second free parameter determined in the fits is a possible
%a global shift of all MC expected shower maximum depth relative 
%a to the measured distribution (Figure\,\ref{fitresult}, bottom).
%a INTERPRETATION, VERGLEICH MIT SYSTEMATISCHEM FEHLER. 

%
%a\begin{figure}[ht]
%a\vspace{0cm}
%a\hspace{0cm}\epsfxsize=8.8cm \epsfbox{fitresult.eps}
%a\vspace{0cm}
%a\caption{  
%aTop: the measured fraction (not corrected for the energy reconstruction
%abiases) of light nuclei as a function of energy.\newline
%aMiddle: the ${\rm \chi^2/Dof}$ of the fits to the shower maximum
%adepth distributions in intervals of the reconstructed energy.\newline
%aBottom: shift of the shower maximum depth with respect to the 
%aMC (QGSJET within CORSIKA\,5.20) expectation.
%a}
%a\label{fitresult}
%a\end{figure} 
%
%  
\subsection{Elongation rate}
\label{sec_elon}
Figure\,\ref{hmax} shows the corrected mean shower maximum 
depth as a function of energy.
A least-squares fit to the $X_{\rm max}$ values as a function of energy, using
only the statistical errors, 
\begin{equation}
X_{\rm max} = \rm{ER \times log_{10}(E) + ERB}
\label{eloneq}
\end{equation}
yields an elongation rate
ER=78.3 $\pm$ 1.0 (stat) $\pm$ 6.2 (syst) g/cm$^2$ and mean depth parameter
ERB=243.1 $\pm$ 2.6 (stat) $\pm$ 15.7 (syst) g/cm$^2$. The specified
mean values and statistical errors
are the mean of fit values with the four energy-reconstruction
methods. The systematic error is estimated as the
standard deviation of the mean values inferred with the four
energy-reconstruction methods. The systematic error
introduced by the systematic uncertainty in {\em slope}
is smaller (about 3 and 14 g/cm$^2$ for ER and ERB respectively).  
The reduced $\chi^2$ values of the fit to relation
(\ref{eloneq}) (4 d.o.f.)
are very large (6.6,9.2,17.2,23.5) for energy-reconstruction
methods 1-4, i.e the systematic errors dominate over the rather small
statistical errors for the mean $X_{\rm max}$. Therefore the 
specified estimates of the
statistical errors obtained
with the procedure explained in section \ref{syst} have to be
treated with caution.
The data point at the highest energy lies
about 20 g/cm$^2$ higher in the atmosphere than expected for a
constant elongation rate.
\\
These results are not in contradiction with previous
measurements in this energy range (\cite{compilat}; \cite{turver}). 
This elongation rate, and also the absolute
$X_{\rm max}$, is consistent with data at
higher energies, obtained mainly by the Yakutsk and Fly's
Eye collaboration (\cite{watson}). A constant elongation rate of
$\approx$  73 g/cm$^2$ from 300 TeV up to 10$^7$ TeV
(dotted fit line in the summary diagram 10 in (\cite{watson}))
is an intriguing hypothesis which is
not in contradiction with our data.
\subsection{Fluctuation of shower
penetration depth}
\label{sec_fluc}
\begin{figure}[ht]
\vspace{0cm}
\hspace{0cm}\epsfxsize=8.8cm \epsfbox{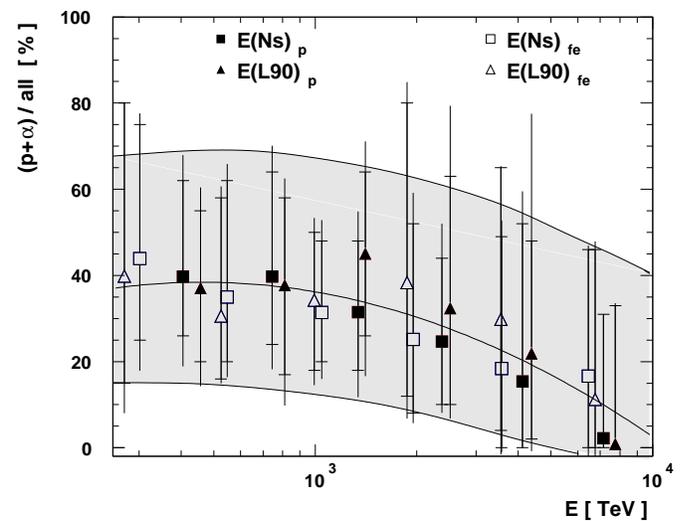}
\vspace{0cm}
\caption{  
The corrected fraction of light nuclei determined with four different 
energy reconstruction methods (see text for details).
Each data point is plotted at the true mean energy
of the events used to infer the mean $X_{\rm max}$.
The error bars are statistical and systematic error
added in quadrature (up to the tick-mark: only statistical error).
The statistical errors are correlated due to the
use of an identical Monte-Carlo sample in the first  
and last three energy intervals.  
The shaded band shows the allowed region between a 
polynomial fit
to the upper and lower ends of the error bars.
}
\label{compos}
\end{figure}

The RMS of the penetration depth
distributions - calculated
in reconstructed-energy bins, i.e.
biased in favour of the light component of CR
especially at low energies -
is shown in Table \ref{table1}. It does not show
any obvious trend towards a heavy composition.
Therefore the fact that the composition at the highest energy 
seems to be heavy with all energy reconstruction
methods (Fig.\,\ref{compos}) is mainly determined  
by the fact that the $X_{\rm max}$ in the highest energy
bin lies about 20 g/cm$^2$ below 
a constant elongation rate.

\section{Further studies of systematic uncertainties;
analysis methods independent of absolute $X_{\rm max}$}
\label{sec_reli}
The trend for an enrichment in heavy elements 
above the knee - which Fig.\ref{compos}
suggest - is not significant within our errors.
In this section we elucidate this fact further,
and explore what it would take to detect
significantly a  modest trend for an
enrichment in heavy elements - as expected e.g. in a diffusion
model of the knee (see Introduction) -
with the present techniques.
\begin{figure}[ht]
\vspace{0cm}
\hspace{0cm}\epsfxsize=8.8cm \epsfbox{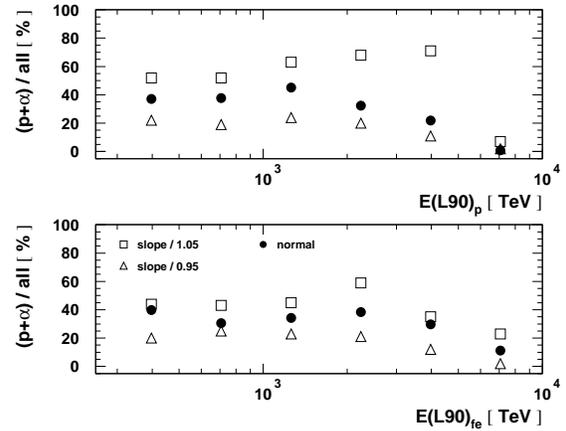}
\vspace{0cm}
\caption
{  
The inferred chemical composition using
energy-reconstruction method 3 (upper panel)
and 4 (lower panel). The dots are the values as discussed before,
the squares (triangles) are the results obtained when the
{\it slope} is increased (decreased) by 5 $\%$ (the systematic error
on this variable).
The general ``trend'' (composition gets heavier/lighter) 
may change within this systematics. 
}
\label{systl90}
\end{figure}
% To derive it we assumed
%that the Monte Carlo used is reliable and that there
%are no energy dependent systematic errors.
%rWe will see that this conclusion rests
%rmainly upon the deviation of the X$_{\rm max}$ values 
%rat the highest energies by up to about 20 g/cm$^2$ below
%rthe one expected from a constant elongation rate.The reasonable
% agreement between the detailed
The agreement between
the $X_{\rm max}$ distribution shape at low energies - predicted assuming
a composition at low energies which is not very different
from the one obtained by direct experiments - and the
data (see Fig.\ref{fitmax}) is satisfactory. This is an argument
in favour of a correct MC simulation.
\\
To explore the effect of our systematic
error in {\it slope} (section \ref{syst}) 
Fig.\ref{systl90} shows the results derived
with an initial assumption of {\it slope}
changed by 5 \% from its preferred value for
energy-reconstruction method 3 and 4.
It becomes clear that not only the mean but even
the overall apparent 
``trend'' may change, e.g. for method 3 with {\it slope} 
increased by  5 \% the composition appears to become {\it lighter}
from the knee up to the penultimate bin.
It should be stressed that none of the discussed ``trends''
is significant within our statistical errors.
\\
The deviation of the penetration depth at the highest
energy point from a constant elongation rate 
discussed in subsection \ref{sec_fluc}
is of the order of disagreements between different Monte Carlo
codes at this energy(\cite{cors2}).
Therefore the possibility remains open that
this deviation is due to a change in cascade characteristics
not reliably modelled by the Monte Carlo, rather than to an enrichment in
heavy elements. As it is also quite
similar in size to our systematic error
in slope, an origin in the HEGRA experiment for this
deviation is also difficult to rule out.
%The above mentioned deviation amounts to about 5 $\%$ 
%in the parameter slope, on the order of our systematic error
%in this parameter (section \ref{syst})
\begin{figure}[ht]
\vspace{0cm}
\hspace{0cm}\epsfxsize=8.8cm \epsfbox{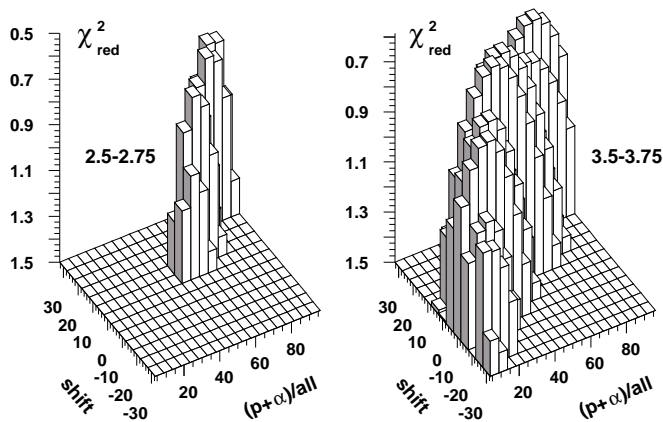}
\vspace{0cm}
\caption{  
The reduced $\chi^2$ values (z-axis) 
of a description of the measured penetration-
depth distribution with the spectral Monte Carlo data as a function
of ``fraction of light nuclei'' (y-axis) and overall shift
in depth (x-axis). Only 
acceptable $\chi^2$ values smaller than 1.5 are
displayed.
%y-axis legend:  ``fraction light''
%x axis legend:  ``shift in depth [g/cm2]''. 
The energy was reconstructed assuming protons and using
the the Cherenkov light density (method 3).
The left panel is for the first energy bin
(log$_{10}$E[TeV]=2.5-2.75, 19000 data events, 1460 Monte-Carlo events)
the right panel for the penultimate one
(log$_{10}$E[TeV]=3.5-3.75, 369 data events, 98 Monte-Carlo events).
%r There are two distinct local minima in the first bin which both
%r have acceptable $\chi^2$ values with a difference of about
%r 20$\%$ absolute in the fraction of light.
In the high-energy bin practically all chemical compositions
are allowed for certain ``shifts''.
}
\label{d2chi}
\end{figure}
\\
Previous conference publications (\cite{rainer95}; 
\cite{durban}; \cite{alcala})
are superseded by the present results
- differences are mainly due to a more sophisticated
amplifier calibration and the simpler energy reconstruction
for the present analysis.
\begin{table*}[ht]
\vspace{-10pt}
\caption{
The RMS of the penetration depth distributions [g/cm$^2$]
as a function of reconstructed energy 
(given in the same units as in Table 1) in the data and
spectral Monte-Carlo sample.
Given is the value inferred
for the energy bins as defined with energy-reconstruction method 3,
i.e. the specified values contain an A dependent bias.
The first error is statistical and the second systematic (due
to the systematic error in slope).
For the numbers from Monte-Carlo simulations only a statistical
error is given.
``Mixed composition'' represents the expectation 
for our best-fit chemical composition.
%A systematic error was inferred from
%the spread of the central value with the four energy-reconstruction
%methods (an additional overall
%systematic error of about 5 $\%$ due to the systematic slope
%error is not included).
Based on numerical experiments
the statistical error was taken as the inferred
value divided by $\sqrt{\rm number\, of\, events\, in\, bin}$ rather
than half of that, as would be correct for Gaussians, due
to the non-Gaussian tails of the distribution.
The comparison between 
experimental data and Monte Carlo simulations shows no trend
towards a heavy composition at the higher energies.}
%other table ratio of (p+$\alpha$)/all, and correction factors
%other table for energy reconstruction method 3.
\label{table1}
\begin{center}
\begin{tabular}{lcccc}
\hline\hline
Rec. energy & Data & MC:mixed comp. & MC:p & MC:Fe \\
\hline
2.5 - 2.75 (ca. 0.32 - 0.56 PeV) &   84 $\pm$ 0.6 $\pm$ 1.8 & 77 $\pm$ 2 & 
 83 $\pm$ 4 & 50 $\pm$ 3  \\
2.75 - 3. (ca. 0.56 - 1 PeV) &  80 $\pm$ 0.9 $\pm$ 1.4 &  &  & \\
3. - 3.25 (ca. 1 - 1.79 PeV) & 80 $\pm$ 2 $\pm$ 1.0 &  &  & \\
3.25 - 3.5 (ca. 1.79 -3.16 PeV) & 73 $\pm$ 2 $\pm$ 1.1 &  &  & \\
3.5 - 3.75 (ca. 3.16 - 5.62 PeV) & 67 $\pm$ 3 $\pm$ 1.5 & 45 $\pm$ 5  & 
73 $\pm$ 17 & 48 $\pm$ 11 \\
3.75 - 4. (ca. 5.62 - 10 PeV) & 67 $\pm$ 7 $\pm$ 0.9 &  &  & \\
\hline
\hline
\end{tabular}
\end{center}
\end{table*}
In the last two of these publications, we tried to
lessen the dependence of our composition result on the correct
absolute $X_{\rm max}$ values by allowing a free ``shift parameter'';
the fit to the $X_{\rm max}$ was then performed with two free
parameters: the ratio of light to all nuclei and an overall
shift in penetration depth of all MC distributions.
In this way the result is mainly determined by the {\it shape} 
of the $X_{\rm max}$ distributions (in first order its
width, i.e. RMS value).
This width depends only weakly on a systematic uncertainty
in the determination
of {\em slope}, relative to the expected difference of a purely light
or heavy composition. 
%r Moreover, as demonstrated 
%r in table 2, the width agrees well among various MCs.
\\
Fig.\ref{d2chi} displays the result of such an attempt 
in a two dimensional plot showing the reduced $\chi^2$ for 
various ``fraction of light nuclei'' - ``shift parameter''
combinations for the data lowest and highest in energy.
The shift is varied in an interval $\pm$ 30 g/cm$^2$,
estimated from the likely systematic uncertainty of our detector
and the Monte Carlo code.
%A problem is immediately visible: there is more than one local
%minimum with acceptable $\chi^2$ values in both cases.
While in the low energy bin small  
(p+$\alpha$)/all ratios lead to unsatisfactory $\chi^2$ values
for all shift values,
in the highest energy bin - well above the knee - practically
all fractions give acceptable $\chi^2$ values for appropriate shifts
as seen in Fig.\ref{d2chi}.
The reason for this behaviour is that 
- given the small number of events in the high-energy bin -
the $X_{\rm max}$
distribution can be fitted both with the relatively broad
predominantly light composition shifted to larger depths in the atmosphere
and a mixed heavy/ light composition (where the difference in mean
penetration depth of the heavy and light component contributes
to the total width) shifted to small penetration depth.
We have to conclude that it is not reliably
possible to determine the composition based mainly on the
width of the $X_{\rm max}$ distribution.
We found in numerical experiments that with this
method, and
assuming a Monte-Carlo simulation describing
the experimental data well, together with 
a statistics increased by about a factor of 100 (which is difficult
but not impossible to reach in future experiments)
it will just be possible
to reach the desired precision of 10 \%
mentioned in the introduction on a 2$\sigma$ level
beyond the knee.
\\
%rWe therefore have to conclude that our results
%ron composition depend heavily on assumptions whose
%rvalidity cannot be chacked reliably at the moment
%rand should not be used to draw physics conclusions.
%rFor further significant progress along the lines of the present
%rpaper, an array reaching systematic errors in slope
%rlow 1 $\%$ and a detection area about 10 times
%rlarger than AIROBICC wuld be needed. 
% 
%__________________________________________________________________

\section{Conclusion}
\label{sec_conc}
The results of this paper demonstrate 
that our data seem to be 
well described by the QGSJET hadron generator within
the CORSIKA program: The energy spectra derived 
with different air-shower observables and assumptions
about the chemical composition of CR
differ only within the estimated systematic errors.
This is an argument in favour of the contention that the ``knee''
is a feature in the primary flux of cosmic rays, rather than
some new effect in the interaction of cosmic rays at very high
energies. Such a new effect 
would probably lead to inconsistencies
in the data analysed by a generator that does not
take them into account.
\\
Up to the knee 
we find a mixed composition and an energy-spectrum power law index
consistent with the results of direct experiments at energies
around 100 TeV.
The data favour a gradual enrichment in heavy elements at energies above the
knee but the systematic errors of our
experiment do not allow to 
rule out a constant composition.
An abrupt change of composition or a substantial enrichment in light
elements at the knee is ruled out.
We find an elongation
rate of about 78 g/cm$^2$ and a mean penetration depth
consistent with other experimental
data on CR with energies higher than studied here (\cite{watson}).
%__________________________________________________________________
\begin{acknowledgements}
We thank the Instituto de Astrofisica de Canarias (IAC) and the 
ORM observatory for support and excellent working conditions.
We are grateful to the authors of CORSIKA and to the 
KASCADE collaboration for many stimulating discussions.\\
The experiment HEGRA is supported by the German BMBF and 
DFG and the Spanish CICYT.  
\end{acknowledgements}

\end{document}